\documentclass[twocolumn,floats,floatfix,showpacs,amssymb,prd,superscriptaddress,nofootinbib]{revtex4-1}
\usepackage{graphicx, epsfig, amssymb} 
\usepackage{amsmath, amsfonts}
\usepackage{bm} 

\usepackage[linktocpage]{hyperref}
\usepackage[caption=false]{subfig}
\usepackage[usenames]{color}

\def\be{\begin{equation}}
\def\ee{\end{equation}}
\def\beq{\begin{eqnarray}}
\def\eeq{\end{eqnarray}}

\begin{document}

\title{\large Head-on collisions of unequal mass black holes in $D=5$ dimensions}

\author{Helvi Witek}\email{helvi.witek@ist.utl.pt}
\affiliation{CENTRA, Departamento de F\'{\i}sica, 
Instituto Superior T\'ecnico, Universidade T\'ecnica de Lisboa (UTL),
Avenida Rovisco Pais 1, 1049 Lisboa, Portugal}

\author{Vitor Cardoso}
\affiliation{CENTRA, Departamento de F\'{\i}sica, 
Instituto Superior T\'ecnico, Universidade T\'ecnica de Lisboa (UTL),
Avenida Rovisco Pais 1, 1049 Lisboa, Portugal}
\affiliation{Department of Physics and Astronomy, 
The University of Mississippi, University, Mississippi 38677, USA}

\author{Leonardo Gualtieri}
\affiliation{Dipartimento di Fisica, ``Sapienza'' Universit\`a di Roma
e Sezione INFN Roma1, P.A. Moro 5, 00185, Roma, Italy}

\author{Carlos Herdeiro}
\affiliation{Departamento de F\'\i sica da Universidade de Aveiro, Campus de Santiago, 3810-183 Aveiro, Portugal}
\affiliation{Centro de F\'\i sica do Porto (CFP),
Departamento de F\'\i sica e Astronomia,
Faculdade de Ci\^encias da Universidade do Porto (FCUP),
Rua do Campo Alegre, 4169-007 Porto, Portugal}

\author{Ulrich Sperhake}
\affiliation{Department of Physics and Astronomy, 
The University of Mississippi, University, Mississippi 38677, USA.}
\affiliation{Institut de Ci\`encies de l'Espai (CSIC-IEEC), 
Facultat de Ci\`encies, Campus Universitat Autonoma de Barcelona, 
E-08193 Bellaterra, Spain}
\affiliation{California Institute of Technology, Pasadena, 
California 91125, USA}

\author{Miguel Zilh\~ao}
\affiliation{Centro de F\'\i sica do Porto (CFP),
Departamento de F\'\i sica e Astronomia,
Faculdade de Ci\^encias da Universidade do Porto (FCUP),
Rua do Campo Alegre, 4169-007 Porto, Portugal}

\begin{abstract}
We study head-on collisions of unequal mass black hole binaries in $D=5$
spacetime dimensions, with mass ratios between 1:1 and 1:4.  Information
about gravitational radiation is extracted by using the Kodama-Ishibashi
gauge-invariant formalism and details of the apparent horizon of the
final black hole. We present waveforms, total integrated energy and momentum for this process.  Our results show
surprisingly good agreement, within $5~\%$ or less, with those
extrapolated from linearized,
point-particle calculations. Our results also show that consistency with
the area theorem bound requires that the same process in a large number
of spacetime dimensions must display new features.
\end{abstract}


\maketitle

\section{Introduction}
Black holes (BHs) have been at the center stage of fundamental physics
in the last decades: supermassive, astrophysical BHs lurk at the center
of most galaxies while large numbers of stellar-mass BHs are thought to
populate each galaxy \cite{Rees1984,Ferrarese:2004qr}; highly dynamical
BH binaries are strong sources of gravitational waves and, perhaps,
power jets and other extreme phenomena \cite{Palenzuela:2010nf,Zanotti:2010xs}. In
high-energy physics, BHs are a central piece of the gauge-gravity
duality \cite{Maldacena:1997re}, and are the generic outcome of
particle collisions at center-of-mass energies above the Planck scale
\cite{Choptuik:2009ww}.  In this regime the particular nature of the
particles' structure should become irrelevant, as indicated by Thorne's
``hoop'' conjecture \cite{Thorne:1972ji}, and ``no-hair theorem''-type of
arguments. These can be invoked to argue that, in general, trans-Planckian
collisions of particles are well described by collisions of highly
boosted BHs. In this context, scenarios such as TeV-gravity are
especially interesting, as they lower the Planck scale to the level
at which BHs would be produced in cosmic rays and particle accelerators
\cite{Banks:1999gd,Giddings:2001bu,Dimopoulos:2001hw,Ahn:2002mj,
Ahn:2003qn,Feng:2001ib, Cardoso:2004zi,
Cavaglia:2002si,Kanti:2004nr,Solodukhin:2002ui}.  Thus, high-energy
BH collisions could be used to look for signatures of extra dimensions
and BH production in ground-based experiments in the forthcoming years.
At the fundamental level, BHs might hold the key for a theory of quantum
gravity, and might help understand important issues such as cosmic
censorship, information loss and the maximum possible luminosities in
any physical process \cite{Sperhake:2008ga,Sperhake:2009jz}.

The above arguments illustrate the necessity to understand accurately
dynamical BH spacetimes, and their potential across a wide variety of
fields. Since the full system of Einstein equations needs to be carefully understood, this is by
all means a monumental task, and typically requires numerical methods.
With these fundamental issues as motivation, long-term efforts to
understand dynamical BHs in generic spacetimes have been initiated
\cite{Yoshino:2009xp,Zilhao:2010sr,Witek:2010xi,Sorkin:2009bc,Sorkin:2009wh,
Dennison:2010wd},
ranging from the inspiralling of BH binaries \cite{Pretorius:2007nq,Hinder:2010vn,Centrella:2010mx},
high-energy collisions of BHs in four
\cite{Sperhake:2008ga,Shibata:2008rq,Sperhake:2009jz} and low energy
collisions in higher spacetime dimensions
\cite{Zilhao:2010sr,Witek:2010xi}, stability studies in higher dimensions
\cite{Shibata:2009ad, Shibata:2010wz,Lehner:2010pn} and BH evolutions
in non asymptotically flat spacetimes \cite{Witek:2010qc}.

Our group has recently studied head-on collisions of equal-mass black
holes in higher dimensions, in particular $D=5$ \cite{Zilhao:2010sr,Witek:2010xi}, 
hereafter denoted as Paper I and Paper II, respectively.
In the present work, we wish to extend that study to the case of
unequal mass BH binaries.  This is an interesting extension for several
reasons, perhaps the most important of which is the nontrivial comparison
with point-particle (PP) calculations in the linearized regime. 
We will compare radiated energy, momentum and multipolar dependence
of our full nonlinear results with results from linearized Einstein
equations. It turns out that the agreement is remarkable, providing
an outstanding consistency check on our codes and results.  A thorough
analysis of the linearized Einstein equations is done in an accompanying
paper \cite{Berti:2010gx}.

This paper is organized as follows: In Sec.~\ref{sec:numres} we summarize 
our numerical method and setup and present the numerical results.
We finish by giving some conclusions and final 
remarks in Sec.~\ref{sec:conclusion}.

\section{Numerical results}
\label{sec:numres}
\begin{table*}
\begin{tabular}{c|cccccc}
\hline
Run & $q$ & $r^{D-3}_{S,1}/r^{D-3}_S$ 
& $r^{D-3}_S$ & $z_1/r_S $ & $z_2 / r_S$ &
Grid Setup \\ 
\hline
HD5a & $1$ & $0.5$ 
& $1$ & $3.185$ & $-3.185$ 
& $\{(256,128,64,32,16,8)\times(2,1,0.5),~h=1/84\}$ \\
HD5b & $1/2$ & $0.33$ 
& $1.5$ & $4.247$ & $-2.123$ 
& $\{(209,104.5,52.3,26.1,13.1,6.5)\times(1.6,0.8,0.4),~h=1/102.9\}$ \\
HD5c & $1/3$ & $0.25$ 
& $2$ & $4.777$ & $-1.592$ 
& $\{(181.0,90.5,45.3,22.6,11.3)\times(2.8,1.4,0.7,0.4),~h=1/118.8\}$ \\
\hline
HD5d$_c$ & $1/4$ & $0.2$ 
& $2.5$ & $5.096$ & $-1.274$ 
& $\{(161.9,80.9,40.5,20.2,10.1)\times(2.5,1.3,0.6,0.3),~h=1/113.8\}$ \\
HD5d$_m$ & $1/4$ & $0.2$ 
& $2.5$ & $5.096$ & $-1.274$ 
& $\{(161.9,80.9,40.5,20.2,10.1)\times(2.5,1.3,0.6,0.3),~h=1/123.3\}$ \\
HD5d$_f$ & $1/4$ & $0.2$ 
& $2.5$ & $5.096$ & $-1.274$ 
& $\{(161.9,80.9,40.5,20.2,10.1)\times(2.5,1.3,0.6,0.3),~h=1/132.8\}$ \\
\hline
\end{tabular}
\caption{\label{tab:setup}
  Grid structure and initial parameters of the head-on collisions from rest in $D=5$.
  The grid setup is given in terms of the ``radii'' of the individual refinement levels,
  as well as the resolution near
  the punctures $h$, 
  in units of $r_S$ (see Sec.~II E in \cite{Sperhake:2006cy} for details).
  We give the Schwarzschild radius $r^{D-3}_{S} = r^{D-3}_{S,1} + r^{D-3}_{S,2}$ 
  of the final BH,
  the mass of the smaller BH $r^{D-3}_{S,1} / r^{D-3}_{S}$ 
  and the quantity $q \equiv M_1/M_2= r^{D-3}_{S,1}/r^{D-3}_{S,2}$ 
  denotes the mass ratio.
  $z_{i}/r_S$ 
  is the initial position of $i$-th BH.
}
\end{table*}
The numerical simulations have been performed with the \textsc{Lean} code, 
originally introduced in Refs.~\cite{Sperhake:2006cy,Sperhake:2007gu}
and adapted to higher dimensional spacetimes in Paper I.
The \textsc{Lean} code is based on the \texttt{Cactus}
computational toolkit \cite{cactus} and uses the \texttt{Carpet} mesh refinement package 
\cite{Schnetter:2003rb, carpet} and Thornburg's apparent horizon finder 
\texttt{AHFinderDirect} \cite{Thornburg:1995cp,Thornburg:2003sf}. 

Following the approach developed in Papers I and II,
we perform a dimensional reduction by isometry of a $D$-dimensional ($D\ge5$), 
asymptotically flat spacetime with $SO(D-2)$ isometry group; 
this is the symmetry, for instance, of a head-on collision of black holes in $D$ dimensions. As discussed in 
Paper I,
the dimensional reduction is performed on the $(D-4)$-sphere $S^{D-4}$, 
which is the symmetry manifold generated by the subgroup $SO(D-3)\subset SO(D-2)$. We remark that the dimensional
reduction is not a compactification, but simply a way to employ the symmetries of the problem in order to rewrite the $D$~dimensional vacuum Einstein equations
as an effective $3+1$ dimensional time evolution problem with source terms that involve a scalar field
(see Eqs.~(2.16)-(2.18) in Paper I).

The coordinate frame in which the numerical simulations are performed is
\begin{equation}
(x^\mu,\phi^1,\dots,\phi^{D-4})=
(t,x,y,z,\phi^1,\dots,\phi^{D-4}) \, ,
\label{frame1}
\end{equation}
where the angles $\phi^1,\dots,\phi^{D-4}$ describe the quotient manifold $S^{D-4}$ 
and do not appear explicitly in the simulations. Here, $z$ is the symmetry axis, i.e.\ the collision line.

We have evolved this 
system using the Baumgarte-Shapiro-Shibata-Nakamura
\cite{Shibata:1995we, Baumgarte:1998te} formulation along the
lines presented in Paper I
(see eqs. (2.43a)-(2.49b) therein), together with the moving
puncture approach \cite{Baker:2005vv,Campanelli:2005dd}. The initial
data consist in the time-symmetric Brill-Lindquist initial data in the
form presented in Paper II (Eq.~(2.15) therein).
Gravitational waves have been extracted using the Kodama-Ishibashi (KI)
formalism \cite{Kodama:2000fa,Kodama:2003jz}. For details of the wave
extraction implementation we refer the reader to Paper II.
We have evolved BH binaries, colliding head-on from rest
with mass ratios
$q \equiv M_1/M_2 = r^{D-3}_{S,1} / r^{D-3}_{S,2} = 1,\,1/2,\,1/3,\,1/4$,
where $M_i$ is the mass of the $i$-th BH.  
The mass parameter $r^{D-3}_{S,1}/ r^{D-3}_S$ of the smaller BH is
given in Table~\ref{tab:setup} and we adapt the value of the second BH 
accordingly.
The initial coordinate separation of the two BHs is set to $d/r_S =
6.37$ which translates to a proper initial separation of $L/r_S = 6.33$.
Further details of the setup of the simulations are summarized in
Table~\ref{tab:setup}.  Unless denoted otherwise, our discussion will
always refer to the highest resolution runs with 
$h_f/r_S = 1/84$, $h_f/r_S = 1/102.9$, $h_f/r_S = 1/118.8$ and 
$h_f/r_S = 1/132.8$ for models HD5a, HD5b, HD5c and HD5d$_f$ 
in Table~\ref{tab:setup}, respectively.
The energy flux is computed according to Eq.~(2.56) in
Paper II (see Eq.~(21) in Ref.~\cite{Berti:2003si} for
the corresponding expression in Fourier space). The momentum flux can
be obtained from
\be
\frac{dP^i}{dt}=\int_{S_{\infty}}d\Omega \frac{d^2E}{dt d\Omega}n^i\,,
\ee
with $n^i$ a unit radial vector on the sphere at infinity $S_{\infty}$. 
This results in an infinite series coupling different multipoles. 
Using only the first two terms in the series, we find, for instance,
that in $D=5$ the momentum flux in the collision direction is given by
\begin{align}
\label{eq:5Dmomentum}
\frac{dP}{dt} & = \frac{1}{4 \pi} \Phi^{l=3}_{,t}
                   \left( 5 \Phi^{l=2}_{,t} +21 \Phi^{l=4}_{,t} \right) \,.
\end{align}
Here, $\Phi^l_{,t}$ is the $l-$pole component of the KI gauge-invariant wavefunction
\cite{Kodama:2000fa,Kodama:2003jz,Witek:2010xi}.
From the momentum radiated, the recoil velocity of the system can be obtained as
\be
\label{eq:kickvelocity}
v_{\rm recoil} = \left| \int^{\infty}_{-\infty} dt \frac{d P}{d t} \right|\,.
\ee
%
\subsection{Waveforms}
%
\begin{figure}
\begin{center}
\includegraphics[width=0.47\textwidth]{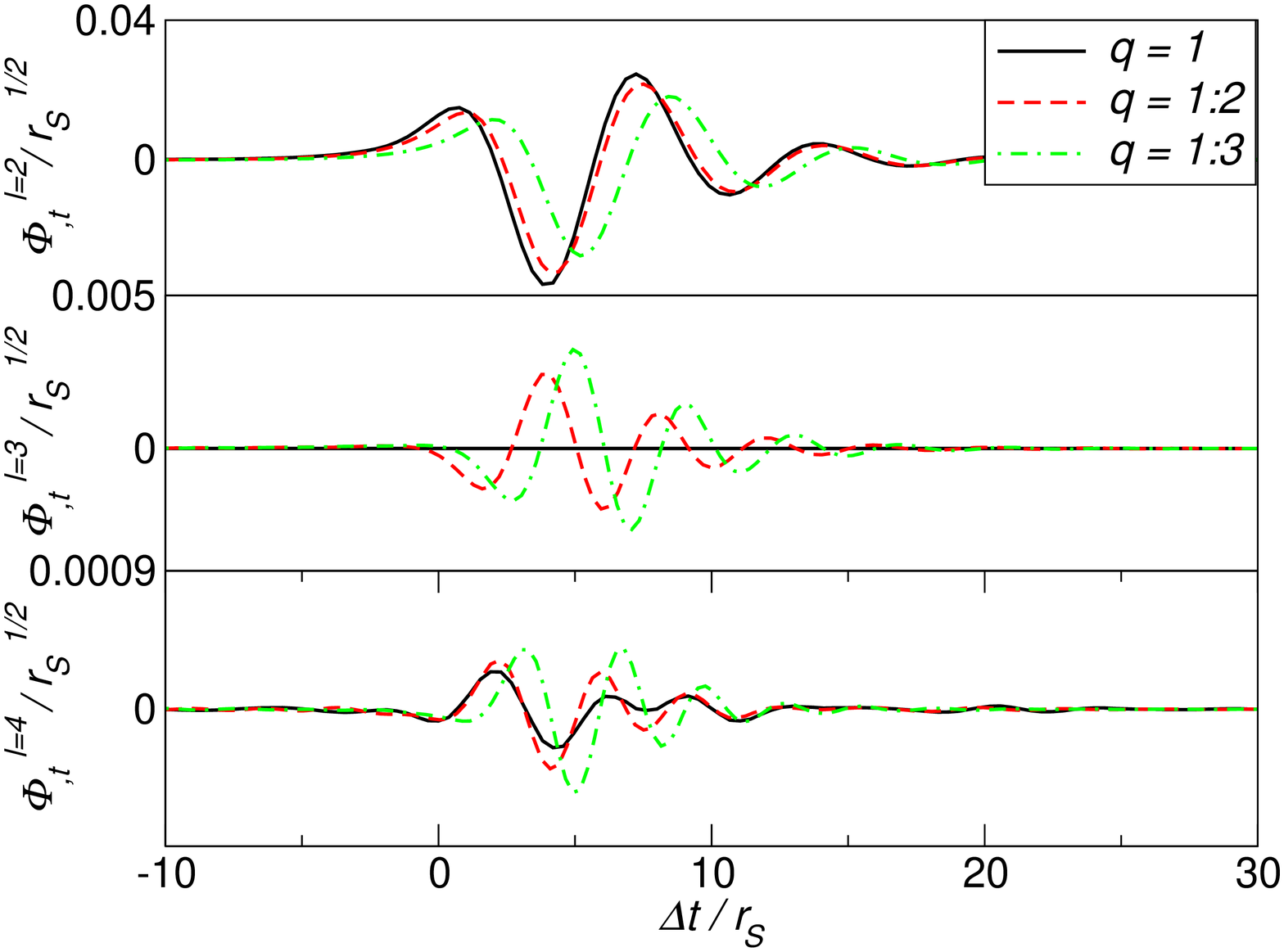}
\end{center}
\caption{\label{fig:waveform} (Color online)
From top to bottom: $l=2$, $l=3$ and $l=4$ modes of the KI waveform
for the different mass ratios; $q = 1$ (black solid lines), $q = 1/2$
(red dashed lines) and $q = 1/3$ (green dash-dotted lines).  The curves
have been shifted in time such that the formation of the common apparent
horizon corresponds to $\Delta t / r_S = 0$ and taking into account the
time that it takes the
waves to propagate to the extraction radius.
}
\end{figure}
In Fig.~\ref{fig:waveform} we show the $l=2,3,4$ waveforms for
different mass ratios, zoomed in around the time of the merger. 
The waveforms have been shifted in time such that
$\Delta t / r_S = (t - r_{ex} - t_{CAH})/r_S = 0$ corresponds to the
time $t_{CAH}$ at which the common apparent horizon forms and taking
into account the propagation time of the waves to the extraction
radius $r_{ex}/r_S = 60, 49, 42.4, 37.9$.  
The waveform is similar to previous four-dimensional
results (see, e.g. Ref.~\cite{Anninos:1998wt}, a more detailed study is in preparation \cite{D4UnequalMass}). 
Although not shown in Fig.~\ref{fig:waveform} we observe a small, spurious signal starting
around $(t - r_{ex})/r_S \approx 0$, which is an artifact of the initial data.

The actual physical part of the waveform is dominated by the merger signal
at $\Delta t / r_S \approx 0$ followed by the quasinormal ringdown.
We estimate that the different ringdown modes are given by
\beq
\omega_{l=2} r_S&=&0.955 \pm 0.005 - i (0.255 \pm 0.005)\,,\nonumber\\
\omega_{l=3} r_S&=&1.60 \pm 0.01 - i (0.31 \pm 0.01)\,,\nonumber\\
\omega_{l=4} r_S&=&2.25 \pm 0.03 - i (0.35 \pm 0.05)\,.\nonumber
\eeq
These results agree well, and within uncertainties, with estimates from
linearized theory
\cite{Berti:2009kk,Cardoso:2003qd,Konoplya:2003dd,Berti:2003si}, providing
a strong
consistency check on our results.
\begin{figure}
\begin{center}
\begin{tabular}{c}
\includegraphics[width=0.47\textwidth]{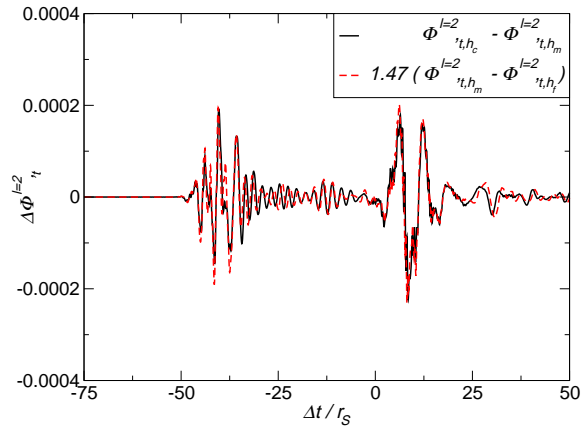} 
\end{tabular}
\end{center}
\caption{\label{fig:convergence} (Color online)
Convergence analysis of the $l=2$ mode
for model ${\rm HD5d}$ in Table~\ref{tab:setup}. We show the differences
between the coarse and medium resolution waveform (black solid line)
and the medium and high resolution waveform (red dashed line). The
latter has been amplified by the factor $Q=1.47$, indicating fourth
order convergence.  The curves have been shifted in time such that the
formation of the common apparent horizon corresponds to $\Delta t /
r_S = 0$ and taking into account the time that it takes the
waves to propagate to the extraction radius. 
}
\end{figure}
Finally, we consider numerical convergence of our waveforms. This study
is summarized in Fig.~\ref{fig:convergence} for the $l=2$ mode of the KI
wavefunction, and for the most challenging mass ratio, $q=1/4$, model
${\rm HD5d}$ in Table~\ref{tab:setup}.  We have evolved this setup at
three different resolutions, namely 
$h_c/r_S = 1/113.8$, $h_m/r_S = 1/123.3$ and $h_f/r_s = 1/132.8$,
which we will refer to as ``coarse'', ``medium''
and ``high'' resolution in the following.  We show the difference between
the coarse and medium as well as between the medium and high resolution
waveforms. The latter has been amplified by the factor $Q=1.47$, which
indicates fourth order convergence. We obtain the same order of accuracy
for the higher modes. The discretization error in the waveforms is
estimated to be $\approx 1.5\%$.
\subsection{Radiated energy}
%
\begin{table}
\begin{tabular}{c|ccccc}
\hline
$q$      & $ E^{\rm rad}/M (\%)$ & $E^{\rm rad}_{l=2}(\%)$ & $E^{\rm rad}_{l=3}(\%)$ & $E^{\rm rad}_{l=4}(\%)$& $v_{\rm recoil} (km/s)$ \\
\hline
1/1     & $0.089 (0.090)$ & $99.9$ & $0.0$ & $0.1$ & $0.00$  \\
1/2     & $0.073 (0.067)$ & $97.7$ & $2.2$ & $0.1$ & $11.37$ \\
1/3     & $0.054 (0.051)$ & $94.8$ & $4.8$ & $0.4$ & $12.64$ \\
1/4     & $0.040 (0.035)$ & $92.4$ & $7.0$ & $0.6$ & $11.38$ \\
\hline
\end{tabular}
\caption{\label{tab:D5EnergyModes}
Summary of our results concerning unequal mass head-on collisions of
BHs in $D=5$.  We show the total radiated energy $E/M$ as measured from
the energy flux at $r_{ex}$, 
the quantity in parenthesis refers
to the estimate obtained using properties of the apparent horizon (see
Paper II for details).  The next three columns show the
fraction of energy $E^l$ excited in the $l$-th mode as compared to the
total radiated energy. The last column refers to the recoil velocity
$v_{\rm recoil}$ in $km/s$.
}
\end{table}
Table~\ref{tab:D5EnergyModes} lists some of the most important physical
quantities which characterize the head-on collision of BHs in $D=5$.
In particular, we show the radiated energy in units of total mass $M$,
and the recoil velocity of the final BH in ${\rm km/s}$. The maximum
amount of energy is emitted in the equal mass case ($E^{\rm rad}/M
= 0.089\%$ as found previously in Paper II), and it
decreases for smaller mass ratios.  We estimate the error in the radiated
energy to be about $5~\%$.
These results have been obtained by integrating
the energy flux as given by the KI master wavefunction. We have also
estimated the radiated energy using properties of the apparent horizon
as described in Paper II.
We estimate the discretization error to be about $\approx10~\%$ when using
this method.
The apparent horizon estimate for the total radiated
energy is shown in parenthesis in Table~\ref{tab:D5EnergyModes}, and
is consistent with the flux computation
within numerical uncertainties.
\begin{figure}
\begin{center}
\begin{tabular}{c}
\includegraphics[width=0.47\textwidth]{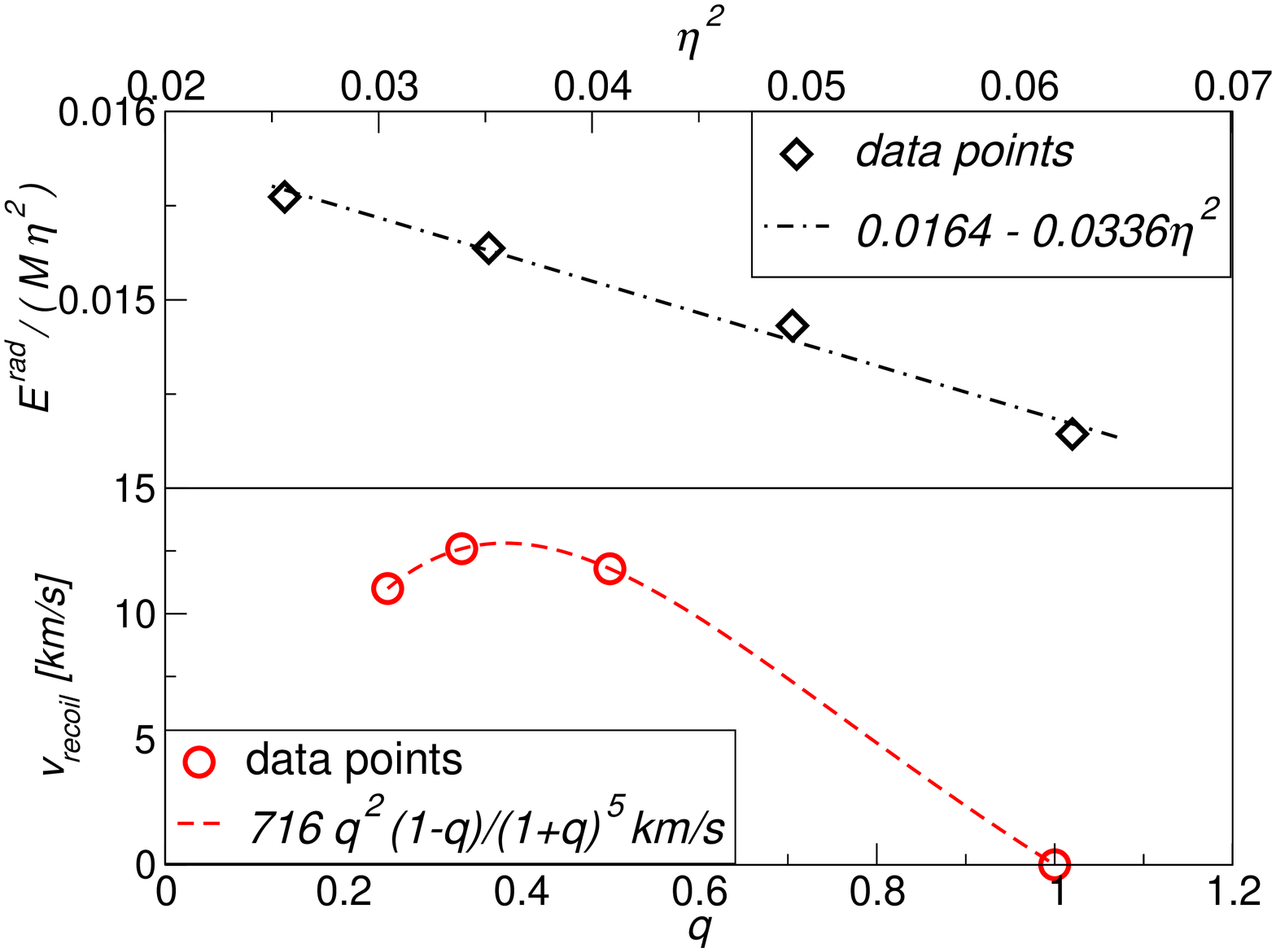}
\end{tabular}
\end{center}
\caption{\label{fig:integratedenergy} (Color online)
Top:
Total integrated energy for different mass ratios, as function of
$\eta^2=[q/(1+q)^2]^2$.  The black diamonds denote the numerical data
and the black dashed-dotted line is the corresponding fitting function,
Eq.~(\ref{eq:fitA}).
Bottom:
Recoil velocity $v_{\rm recoil}$ in $km/s$ as function of the mass
ratio $q$.  The red circles denote the numerical data and the red dashed
line is the corresponding fitting function, Eq.~(\ref{eq:fitV}).
}
\end{figure}
Table~\ref{tab:D5EnergyModes} also shows the fraction of energy emitted in 
different multipoles.
Higher multipoles are clearly enhanced as the mass ratio decreases, 
in agreement with what we expect in the extreme case of a PP 
falling into a BH. In fact, we can make this statement more precise. 
Post-Newtonian arguments, which extend to generic $D$-dimensions, allow one 
to expect the functional dependence for the total radiated energy 
\cite{MadalenaThesis}, $E^{\rm rad}/M \propto \eta^2$, where $\eta=q/(1+q)^2$ is the 
dimensionless reduced mass. For clarity, we show the ratio $E^{\rm rad}/(M\eta^2)$
in the top panel of Fig.~\ref{fig:integratedenergy}, which can be seen to 
depend very weakly on $\eta^2$. 
We can refine this argument by fitting our numerical results to an improved 
expression of the form $E^{\rm rad}/M \eta^2= A_0 + A_1 \eta^2$. We find
\be
\label{eq:fitA}
\frac{E^{\rm rad}}{M \eta^2}= 0.0164 -0.0336 \eta^2 \,.
\ee
Moreover, the following expressions for the multipolar content 
provide a good fit to our numerical data,
\be
\frac{E^{\rm rad}_{l=2}}{E^{\rm rad}}= 0.79 +0.83 \eta \,,\quad
\frac{E^{\rm rad}_{l=3}}{E^{\rm rad}}= 0.19 -0.77 \eta \,.
\ee
Linearized, PP calculations presented in an accompanying paper \cite{Berti:2010gx} 
show that in the limit of zero mass ratio one obtains
\be
\frac{E^{\rm rad}_{\rm PP}}{M\eta^2}=0.0165 \,,
\ee
which agrees with the extrapolation of our numerical results within less than $1\%$.
The multipole contents in the PP limit are 
\be
\frac{E^{\rm rad}_{l=2}}{E^{\rm rad}_{\rm PP}}= 0.784  \,,\quad
\frac{E^{\rm rad}_{l=3}}{E^{\rm rad}_{\rm PP}}= 0.167  \,,
\ee
still in very good agreement with the extrapolation of our full numerical results 
to the zero mass ratio limit.
In fact, bearing in mind that we are extrapolating from mass ratios of $1/4$ 
down to the zero mass ratio limit, the agreement is impressive.
Finally, all these results are consistent with the fact that higher
multipoles contribute more to the radiation than in $D=4$, where for
instance the $l=3$ mode contributed roughly $10\%$ of the total energy in
the PP limit \cite{Davis:1971gg}. Linearized, point-particle calculations
show that the trend is consistent and continues in higher-$D$ \cite{Berti:2010gx},
which might mean that accurate wave extraction will become extremely difficult, as higher
resolution is necessary to resolve higher-$l$ modes \cite{Berti:2007fi}.

\subsection{Radiated momentum}
%
\begin{figure}
\begin{center}
\includegraphics[width=0.47\textwidth]{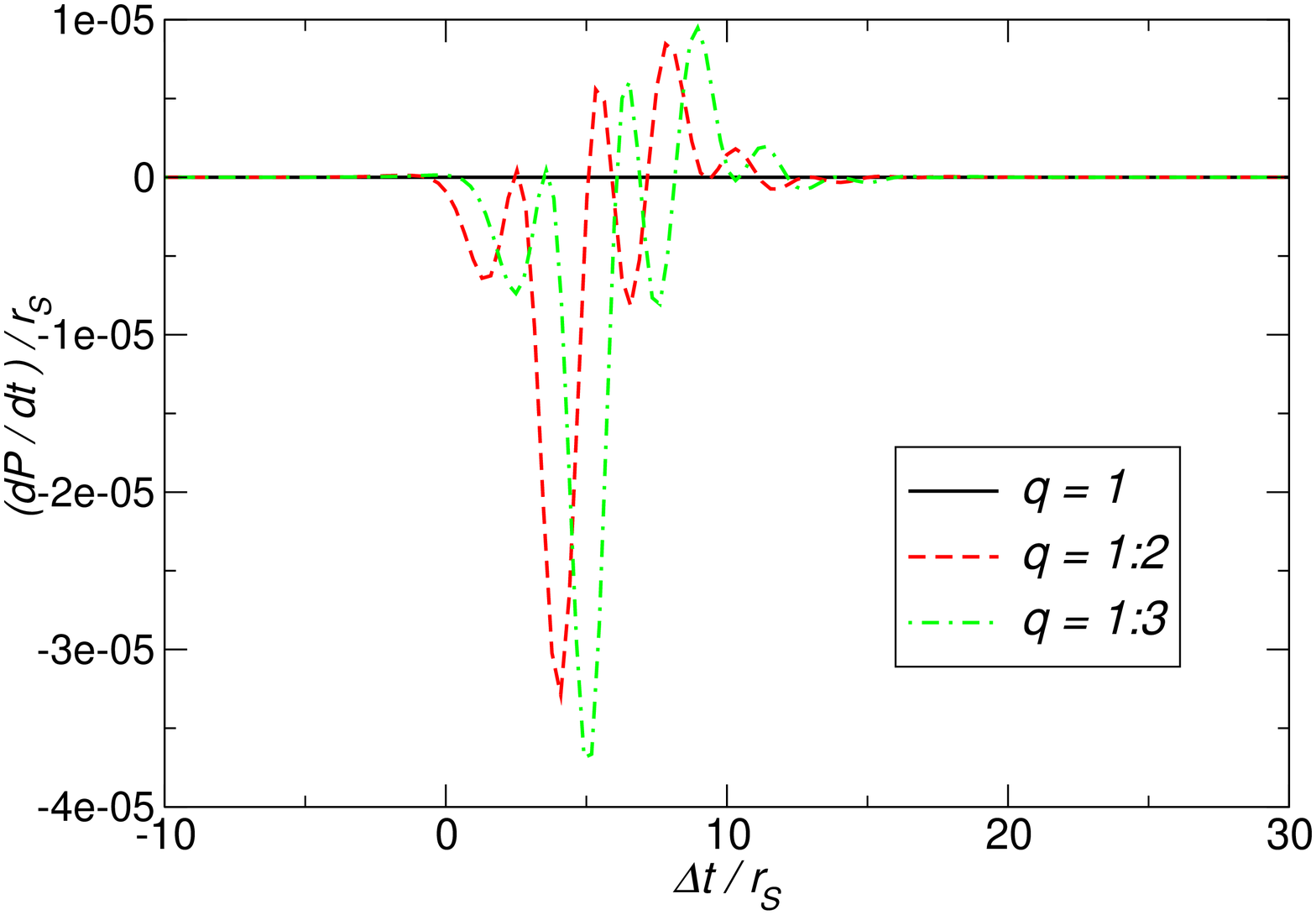} 
\end{center}
\caption{\label{fig:momentumflux} (Color online)
Momentum flux for different mass ratios.
The curves have been shifted in time such that the formation of the common apparent horizon
corresponds to $\Delta t / r_S = 0$ and taking into account the time that it takes the 
waves to propagate to the extraction radius.
}
\end{figure}
For unequal-mass collisions, the asymmetric emission of radiation along the collision axis causes a net momentum to be carried by
gravitational waves. As such, the final BH will ``recoil'', according to 
Eq.~(\ref{eq:kickvelocity}). Momentum fluxes and recoil velocity  
for different mass-ratios are shown in Fig.~\ref{fig:momentumflux} and the bottom panel of Fig.~\ref{fig:integratedenergy}, respectively.
We estimate the errors in the recoil velocity to be $\approx 5~\%$.
The general functional form for the dependence of momentum on the mass
parameters of the individual holes has been worked out by M. Lemos in
generic spacetime dimensions \cite{MadalenaThesis} and is the same
as in four dimensions
\be
\label{eq:fitV}
v_{\rm recoil} = C \frac{q^2(1-q)}{(1+q)^5}\,.
\ee
By fitting this function to our numerical data, we obtain $C=716~{\rm km/s}$.
Observe that $v_{\rm recoil}$ reaches a maximum value at $q=2-\varphi\simeq 0.38$, where $\varphi$ is the golden ratio. The quality of the fit can be seen in the bottom panel of Fig.~\ref{fig:integratedenergy}, 
where we overplot the numerical data points with the fitting function, 
Eq.~(\ref{eq:fitV}). This exercise is interesting because we can again extrapolate our results to the PP limit. In an accompanying paper, Berti et al. \cite{Berti:2010gx} find
\be
v_{\rm recoil} = 779 q^2 {\rm km/s}\,,
\ee
in reasonably good agreement (better than 10$\%$) with our extrapolation. We note that momentum emission is given by a nontrivial
interference between different multipoles, so this is a nontrivial agreement.
\section{Conclusions}
\label{sec:conclusion}

For the first time, we have evolved unequal-mass BH binaries in higher
dimensions, by solving Einstein equations in the full nonlinear regime.
We have focused on head-on collisions in $D=5$ spacetime dimensions.

The gravitational waveforms are similar to the $D=4$ counterparts
\cite{Anninos:1998wt,D4UnequalMass},
and we were able to estimate the ringdown frequencies of the
lowest multipoles. We find good agreement with published values for
the quasinormal frequencies, extracted in a linearized formalism.
When extrapolated to the zero-mass-ratio limit, our results agree with
linearized calculations \cite{Berti:2010gx} at the $\%$ level or better
for the energy and momentum radiated, as well as for the multipolar
dependence.
This outstanding agreement is one of the
main results of this work.

Our findings, supported by linearized analysis, indicate that the higher
multipoles become more important for larger $D$. This will certainly make
wave extraction at sufficiently large $D$ a more demanding task, since
higher resolutions are necessary to resolve these modes.  The momentum
structure is similar to the four-dimensional case, it would be
interesting to understand if other aspects, such as the antikick,
can still be interpreted in like-manner \cite{Rezzolla:2010df}.
Finally, it would be very interesting to perform an exhaustive set
of simulations in higher $D$: our results, together with linearized
analysis \cite{Berti:2010gx}, suggest a qualitative change in radiation
emission for $D \leq 12-13$.
In fact, this change is required
by the fact that Hawking's area theorem forces the total amount of
gravitational radiation to decrease with $D$, at sufficiently large $D$
\cite{Witek:2010xi,Berti:2010gx}. Understanding the mechanism at play
requires extension of our results to arbitrary spacetime dimensions.

\vskip 1cm
\begin{acknowledgments}
  We would like to thank Emanuele Berti and Andrea Nerozzi for useful discussions and suggestions.
  M.Z. and H.W. are funded by FCT through grants
  SFRH/BD/43558/2008 and SFRH/BD/46061/2008 
  U.S. acknowledges support from the Ram\'on y Cajal
  Programme of the Ministry of Education and Science of Spain, NSF grants
  PHY-0601459, PHY-0652995 and the Fairchild Foundation to Caltech.  This
  work was supported by the {\it DyBHo--256667} ERC Starting Grant and by FCT -
  Portugal through projects PTDC/FIS/098025/2008, PTDC/FIS/098032/2008,
  PTDC/CTE-AST/098034/2008, CERN/FP/109306/2009, CERN/FP/109290/2009
  as well as NSF grant PHY-0900735.  This research was supported by
  an allocation through the TeraGrid Advanced Support Program under
  grant PHY-090003 and an allocation by the Centro de Supercomputaci{\'o}n
  de Galicia (CESGA) under project ICTS-2009-40.
  Computations were performed on the TeraGrid cluster
  NICS Kraken, the Milipeia cluster in Coimbra, Magerit
  in Madrid, Finis Terrae and LRZ in Munich.
  The authors thankfully acknowledge the
  computer resources, technical expertise and assistance provided by the
  Barcelona Supercomputing Centre---Centro Nacional de Supercomputaci\'on.
  
\end{acknowledgments}

\bibliographystyle{h-physrev4}
\bibliography{paperUnequalMassHeadon}

\begin{thebibliography}{10}

\bibitem{Rees1984}
M.~Rees,
\newblock Ann. Rev. Astron. Astrophys. {\bf 22}, 471 (1984).

\bibitem{Ferrarese:2004qr}
L.~Ferrarese and H.~Ford,
\newblock Space Sci. Rev. {\bf 116}, 523 (2005), [astro-ph/0411247].

\bibitem{Palenzuela:2010nf}
C.~Palenzuela, L.~Lehner and S.~L. Liebling,
\newblock Science {\bf 329}, 927 (2010), [1005.1067].

\bibitem{Zanotti:2010xs}
O.~Zanotti, L.~Rezzolla, L.~Del~Zanna and C.~Palenzuela,
\newblock Astron. Astrophys. {\bf 523}, A8 (2010), [1002.4185].

\bibitem{Maldacena:1997re}
J.~M. Maldacena,
\newblock Adv. Theor. Math. Phys. {\bf 2}, 231 (1998), [hep-th/9711200].

\bibitem{Choptuik:2009ww}
M.~W. Choptuik and F.~Pretorius,
\newblock Phys. Rev. Lett. {\bf 104}, 111101 (2010), [0908.1780].

\bibitem{Thorne:1972ji}
K.~S. Thorne,
\newblock In *J. R. Klauder, Magic Without Magic*, San Francisco 1972, 231-258.

\bibitem{Banks:1999gd}
T.~Banks and W.~Fischler,
\newblock hep-th/9906038.

\bibitem{Giddings:2001bu}
S.~B. Giddings and S.~D. Thomas,
\newblock Phys. Rev. {\bf D65}, 056010 (2002), [hep-ph/0106219].

\bibitem{Dimopoulos:2001hw}
S.~Dimopoulos and G.~L. Landsberg,
\newblock Phys. Rev. Lett. {\bf 87}, 161602 (2001), [hep-ph/0106295].

\bibitem{Ahn:2002mj}
E.-J. Ahn, M.~Cavaglia and A.~V. Olinto,
\newblock Phys. Lett. {\bf B551}, 1 (2003), [hep-th/0201042].

\bibitem{Ahn:2003qn}
E.-J. Ahn, M.~Ave, M.~Cavaglia and A.~V. Olinto,
\newblock Phys. Rev. {\bf D68}, 043004 (2003), [hep-ph/0306008].

\bibitem{Feng:2001ib}
J.~L. Feng and A.~D. Shapere,
\newblock Phys. Rev. Lett. {\bf 88}, 021303 (2001), [hep-ph/0109106].

\bibitem{Cardoso:2004zi}
V.~Cardoso, M.~C. Espirito~Santo, M.~Paulos, M.~Pimenta and B.~Tome,
\newblock Astropart. Phys. {\bf 22}, 399 (2005), [hep-ph/0405056].

\bibitem{Cavaglia:2002si}
M.~Cavaglia,
\newblock Int. J. Mod. Phys. {\bf A18}, 1843 (2003), [hep-ph/0210296].

\bibitem{Kanti:2004nr}
P.~Kanti,
\newblock Int. J. Mod. Phys. {\bf A19}, 4899 (2004), [hep-ph/0402168].

\bibitem{Solodukhin:2002ui}
S.~N. Solodukhin,
\newblock Phys. Lett. {\bf B533}, 153 (2002), [hep-ph/0201248].

\bibitem{Sperhake:2008ga}
U.~Sperhake, V.~Cardoso, F.~Pretorius, E.~Berti and J.~A. Gonzalez,
\newblock Phys. Rev. Lett. {\bf 101}, 161101 (2008), [0806.1738].

\bibitem{Sperhake:2009jz}
U.~Sperhake {\em et~al.},
\newblock Phys. Rev. Lett. {\bf 103}, 131102 (2009), [0907.1252].

\bibitem{Yoshino:2009xp}
H.~Yoshino and M.~Shibata,
\newblock Phys. Rev. {\bf D80}, 084025 (2009), [0907.2760].

\bibitem{Zilhao:2010sr}
M.~Zilhao {\em et~al.},
\newblock Phys. Rev. {\bf D81}, 084052 (2010), [1001.2302].

\bibitem{Witek:2010xi}
H.~Witek {\em et~al.},
\newblock Phys.Rev. {\bf D82}, 104014 (2010), [1006.3081].

\bibitem{Sorkin:2009bc}
E.~Sorkin and M.~W. Choptuik,
\newblock Gen. Rel. Grav. {\bf 42}, 1239 (2009), [0908.2500].

\bibitem{Sorkin:2009wh}
E.~Sorkin,
\newblock Phys. Rev. {\bf D81}, 084062 (2010), [0911.2011].

\bibitem{Dennison:2010wd}
K.~A. Dennison, J.~P. Wendell, T.~W. Baumgarte and J.~D. Brown,
\newblock Phys. Rev. {\bf D82}, 124057 (2010), [1010.5723].

\bibitem{Pretorius:2007nq}
F.~Pretorius,
\newblock 0710.1338.

\bibitem{Hinder:2010vn}
I.~Hinder,
\newblock Class. Quant. Grav. {\bf 27}, 114004 (2010), [1001.5161].

\bibitem{Centrella:2010mx}
J.~M. Centrella, J.~G. Baker, B.~J. Kelly and J.~R. van Meter,
\newblock Rev. Mod. Phys. {\bf 82}, 3069 (2010),
\newblock arXiv:1010.5260 [gr-qc].

\bibitem{Shibata:2008rq}
M.~Shibata, H.~Okawa and T.~Yamamoto,
\newblock Phys. Rev. {\bf D78}, 101501 (2008), [0810.4735].

\bibitem{Shibata:2009ad}
M.~Shibata and H.~Yoshino,
\newblock Phys. Rev. {\bf D81}, 021501 (2010), [0912.3606].

\bibitem{Shibata:2010wz}
M.~Shibata and H.~Yoshino,
\newblock Phys. Rev. {\bf D81}, 104035 (2010), [1004.4970].

\bibitem{Lehner:2010pn}
L.~Lehner and F.~Pretorius,
\newblock Phys. Rev. Lett. {\bf 105}, 101102 (2010), [1006.5960].

\bibitem{Witek:2010qc}
H.~Witek {\em et~al.},
\newblock Phys. Rev. {\bf D82}, 104037 (2010), [1004.4633].

\bibitem{Berti:2010gx}
E.~Berti, V.~Cardoso and B.~Kipapa,
\newblock 1010.3874.

\bibitem{Sperhake:2006cy}
U.~Sperhake,
\newblock Phys. Rev. {\bf D76}, 104015 (2007), [gr-qc/0606079].

\bibitem{Sperhake:2007gu}
U.~Sperhake {\em et~al.},
\newblock Phys. Rev. {\bf D78}, 064069 (2008), [0710.3823].

\bibitem{cactus}
{Cactus} {Computational} {Toolkit},
\newblock \url{http://www.cactuscode.org/}.

\bibitem{Schnetter:2003rb}
E.~Schnetter, S.~H. Hawley and I.~Hawke,
\newblock Class. Quant. Grav. {\bf 21}, 1465 (2004), [gr-qc/0310042].

\bibitem{carpet}
Mesh refinement with {Carpet},
\newblock \url{http://www.carpetcode.org/}.

\bibitem{Thornburg:1995cp}
J.~Thornburg,
\newblock Phys. Rev. {\bf D54}, 4899 (1996), [gr-qc/9508014].

\bibitem{Thornburg:2003sf}
J.~Thornburg,
\newblock Class. Quant. Grav. {\bf 21}, 743 (2004), [gr-qc/0306056].

\bibitem{Shibata:1995we}
M.~Shibata and T.~Nakamura,
\newblock Phys. Rev. {\bf D52}, 5428 (1995).

\bibitem{Baumgarte:1998te}
T.~W. Baumgarte and S.~L. Shapiro,
\newblock Phys. Rev. {\bf D59}, 024007 (1998), [gr-qc/9810065].

\bibitem{Baker:2005vv}
J.~G. Baker, J.~Centrella, D.-I. Choi, M.~Koppitz and J.~van Meter,
\newblock Phys. Rev. Lett. {\bf 96}, 111102 (2006), [gr-qc/0511103].

\bibitem{Campanelli:2005dd}
M.~Campanelli, C.~O. Lousto, P.~Marronetti and Y.~Zlochower,
\newblock Phys. Rev. Lett. {\bf 96}, 111101 (2006), [gr-qc/0511048].

\bibitem{Kodama:2000fa}
H.~Kodama, A.~Ishibashi and O.~Seto,
\newblock Phys. Rev. {\bf D62}, 064022 (2000), [hep-th/0004160].

\bibitem{Kodama:2003jz}
H.~Kodama and A.~Ishibashi,
\newblock Prog. Theor. Phys. {\bf 110}, 701 (2003), [hep-th/0305147].

\bibitem{Berti:2003si}
E.~Berti, M.~Cavaglia and L.~Gualtieri,
\newblock Phys. Rev. {\bf D69}, 124011 (2004), [hep-th/0309203].

\bibitem{Anninos:1998wt}
P.~Anninos and S.~Brandt,
\newblock Phys. Rev. Lett. {\bf 81}, 508 (1998), [gr-qc/9806031].

\bibitem{D4UnequalMass}
U.~e.~a. Sperhake,
\newblock {in preparation}.

\bibitem{Berti:2009kk}
E.~Berti, V.~Cardoso and A.~O. Starinets,
\newblock Class. Quant. Grav. {\bf 26}, 163001 (2009), [0905.2975].

\bibitem{Cardoso:2003qd}
V.~Cardoso, J.~P.~S. Lemos and S.~Yoshida,
\newblock JHEP {\bf 12}, 041 (2003), [hep-th/0311260].

\bibitem{Konoplya:2003dd}
R.~Konoplya,
\newblock Phys.Rev. {\bf D68}, 124017 (2003), [hep-th/0309030].

\bibitem{MadalenaThesis}
M.~Lemos,
\newblock Approximation methods in the study of gravitational-wave generation:
  From the quadrupole to the zfl,
\newblock Insituto Superior Tecnico, Universidade Tecnica de Lisboa
  (unpublished),
  \url{http://blackholes.ist.utl.pt//fp-content/attachs/thesismadalenalemos.pd%
f}.

\bibitem{Davis:1971gg}
M.~Davis, R.~Ruffini, W.~Press and R.~Price,
\newblock Phys.Rev.Lett. {\bf 27}, 1466 (1971).

\bibitem{Berti:2007fi}
E.~Berti {\em et~al.},
\newblock Phys. Rev. {\bf D76}, 064034 (2007), [gr-qc/0703053].

\bibitem{Rezzolla:2010df}
L.~Rezzolla, R.~P. Macedo and J.~L. Jaramillo,
\newblock Phys. Rev. Lett. {\bf 104}, 221101 (2010), [1003.0873].

\end{thebibliography}

\end{document}